\documentclass[12pt,preprint]{aastex}

\newcommand{\Deg}{${}^{\circ}$}
\newcommand{\Min}{${}^{\prime}$}
\newcommand{\Sec}{${}^{\prime\prime}$}

\shorttitle{The K-metallicity relation and HCG 31}
\shortauthors{Mendes de Oliveira et al.}

\begin{document}

\title{
The K luminosity-metallicity relation for dwarf galaxies and the
tidal dwarf galaxies in the tails of HCG 31
\thanks{Based on observations obtained at the Gemini Observatory,
which is operated by the Association of Universities for Research in
Astronomy, Inc., under a cooperative agreement with the NSF on behalf of
the Gemini partnership: the National Science Foundation (United States),
the Particle Physics and Astronomy Research Council (United Kingdom),
the National Research Council (Canada), CONICYT (Chile), the Australian
Research Council (Australia), CNPq (Brazil) and CONICET (Argentina) --
Observing run ID: GN-2004B-Q-47.}
}

\author{Claudia L. Mendes de Oliveira}
\affil{Departamento de Astronomia, Instituto de Astronomia, Geof\'{\i}sica
e Ci\^encias Atmosf\'ericas da USP, Rua do Mat\~ao 1226, Cidade
Universit\'aria, 05508-090, S\~ao Paulo, Brazil}
\email{oliveira@astro.iag.usp.br}

\author{Sonia Temporin\altaffilmark{2}}\affil{Institut f\"ur Astrophysik,
Leopold-Franzens-Universit\"at Innsbruck, Technikerstra\ss e 25, A-6020
Innsbruck, Austria}
\altaffiltext{2}{INAF - Osservatorio Astronomico di Brera, Via Brera 28,
I-20121 Milano, Italy}
\email{giovanna.temporin@uibk.ac.at}

\author{Eduardo S. Cypriano}\affil{Southern Astrophysics Research
Telescope, Casilla 603, La Serena, Chile and Laborat\'orio Nacional
de Astrof\'{\i}sica, CP 21, 37500-000 Itajub\'a - MG, Brazil}
\email{ecypriano@ctio.noao.edu}

\author{Henri Plana}\affil{Laboratorio de Astrofisica Teorica e
Observacional, Universidade Estadual De Santa Cruz - Brazil}
\email{plana@uesc.br}

\author{Philippe Amram}\affil{Observatoire Astronomique Marseille-Provence \&
Laboratoire d'Astrophysique de Marseille,
2 Place Le Verrier, 13248 Marseille Cedex 04, France}
\email{philippe.amram@oamp.fr}

\author{Laerte Sodr\'e Jr.}
\affil{Departamento de Astronomia, Instituto de Astronomia, Geof\'{\i}sica
e Ci\^encias Atmosf\'ericas da USP, Rua do Mat\~ao 1226, Cidade
Universit\'aria, 05508-090, S\~ao Paulo, Brazil}
\email{laerte@astro.iag.usp.br}

\author{Chantal Balkowski}
\affil{Observatoire de
Paris, GEPI, CNRS and Universit\'e Paris 7, 5 Place Jules Janssen,
F-92195 Meudon Cedex, France}
\email{chantal.balkowski@obspm.fr} 

\begin{abstract}
We determine a $K$-band luminosity-metallicity (K-Z) relation for
dwarf irregular galaxies, over a large range of magnitudes, $-$20.5 $<$
M$_{K}$ $<$ $-$13.5, using a combination of K photometry from either
the 2-micron all sky survey (2MASS) or the recent study of \citet{v05},
and metallicities derived mainly with the T$_e$ method, from several
different studies. We then use this newly-derived relation, together with
published $K_s$ photometry and our new spectra of objects in the field
of HCG 31 to discuss the nature of the possible tidal dwarf galaxies
of this group.  We catalogue a new member of HCG 31, namely ``R",
situated $\sim$40 kpc north of the group center, composed by a ring of
H$\alpha$ knots which coincides with a peak in HI.  This object is a
deviant point in the K-Z relation (it has too high metallicity for its
luminosity) and its projected distance to the parent galaxy and large
gas reservoir makes it one of the most promising tidal dwarf galaxy
candidates of HCG 31, together with object F.  The subsystems A1, E, F,
H and R all have metallicities similar to that of the galaxies A+C and
B, result that is expected in a scenario where those were formed from
material expelled from the central galaxies of HCG 31. While objects A1,
E and H will most probably fall back onto their progenitors, F and R may
survive as tidal dwarf galaxies.  We find that two galaxies of HCG 31,
G and Q, have A+em spectral signatures, and are probably evolving toward
a post-starburst phase.
\end{abstract}


\section{Introduction}
\addtocounter{footnote}{1}
  It has been established for more than a decade that there is a well
defined B luminosity-metallicity (L-Z) relation for dwarf irregular galaxies,
in the sense that the higher the metallicity, the higher the luminosity
of the dwarf galaxy \citep[e.g. ][]{skh89,rm95}, although \citet{hgo98}
claimed that the relationship was much weaker than previously thought.
More recently the relationship was confirmed by \citet{lee03} with
a root mean square of $\sigma$ = 0.175 dex in $\log$(O/H) for the
same sample of dwarf irregular galaxies originally examined by \citet{rm95}, but
with updated distance determinations and metallicity measurements.  
The L-Z relation is also valid for
giant galaxies, with less scatter and with a steeper slope than for
dwarf systems \citep{s05}.

In the  studies of dwarf systems associated with strongly interacting
galaxies, this luminosity-metallicity relation has been used to select
possible candidate tidal dwarf galaxies (TDGs) -- newly born galaxies
formed out of recycled material expelled from the parent galaxies during
interaction.  TDGs stand out from this relation, i.e., they, in general,
do not follow the L-Z relation of dwarf irregular galaxies, { but
instead have an almost constant metallicity between 1/4 and 1/3 of the
solar value.} Examples of B-band L-Z diagrams showing the location of
tidal dwarf galaxies in a few interacting systems can be found in Fig. 6
of \citet{dm98}, in Fig. 3 of \citet{wdf03} and Fig. 17 of \citet{d00}.
Nevertheless, in systems where star formation is widely spread and the B
luminosities of the parent galaxies and of the TDGs may be altered, the
B-band L-Z relation (used in the studies above) may no longer be a useful
tool to select possible TDGs { mainly because the B band is not a good
tracer of the stellar mass when it is highly affected by starbursts.}
This is the case for the Hickson compact group 31, a gas-rich group,
with intense star forming activity \citep[e.g. ][]{ls04}, { dominated
by a pair of interacting dwarf galaxies, A and C,} which clearly have
their B luminosities quite affected by the light and dust involving
newly born stars \citep[in fact HCG 31C has Wolf-Rayet features in its
spectrum; ][]{c91}.  In the course of studying the possible nature of
the various sub-components of this interesting group, we thus felt the
need of compiling from the literature a K-band L-Z relation.  Such a
relation is more useful than the B-band relation because { it is less
affected by a starburst}, it suffers significantly less from absorption
effects and it better characterizes the bulk of the stars (old component)
in the group member galaxies.

This paper is divided as follows.  In Section~\ref{near}, the K-band
L-Z diagram for ``normal'' dwarf irregular galaxies $-$20.5 $<$ M$_{K}$
$<$ $-$13.5 is determined from literature data.  Section~\ref{obs}
describes our new Gemini data, g$^\prime$ and r$^\prime$ 
photometry and medium-resolution spectroscopy of HCG 31, which are
then used to determine the radial velocities, ages and metallicities of
the regions and to plot the K-band L-Z relation for the objects of this
interacting group { (the K photometry for the HCG 31 members
come mostly from \citet{ls04})}.  Finally, in Section~\ref{Discussion}
we discuss the fate of the TDG candidates of HCG 31.

Distance-dependent measurements assume a distance to the group of 54.8 Mpc
(derived by \citet{ls04} from the Hubble law and 
H$_0$=75 km~s$^{-1}$~Mpc$^{-1}$). We
use the identifications for the HCG 31 
objects suggested by \citet{ls04} with
one exception: their region E corresponds to blob E2 in this paper.
We identify objects with single letters (e.g. objects E, F, R) and we
refer to blobs which form an object with letters followed by numbers
(e.g.  blobs E1 and E2 compose object E).

\section{\label{near}The luminosity-metallicity relation}

\subsection{\label{previouswork}Previous works}

A few authors have questioned the existence of the luminosity-metallicity
relation \citep[e.g. ][]{ca93}, specially for gas-rich objects.  
The $B$ magnitude, classically used to investigate the
L-Z relation, is known to be highly affected by
strong and young starbursts with ages below 10 Myr, as well as by dust.  
Therefore, for the
study of objects like those in HCG 31, which are known to contain young
and strong starbursts, the use of near infrared (NIR) magnitudes should be much more
robust and less affected by the onset of starbursts or dust absorption than the
$B$ magnitude. In the recent study of \citet{s05} the L-Z relation is
actually explored also in the NIR regime for a sample of emission-line galaxies
from the KISS survey. However, they present only global fits to all their sample
galaxies, which are mostly concentrated in the high luminosity part of
the diagrams (M$_K \, < \, -21$), whereas the small numbers of dwarf galaxies
in their sample appear to follow a relation with a shallower slope.
Recent efforts directed to test the existence of an L-Z relation
for dwarf irregular galaxies in the H-band, by following an approach
conceived to minimize the effects of uncertainties in distance determinations \citep{sav05}
have led to encouraging results, although still
based on a small number of galaxies.
Here we investigate the L-Z relation in the $K$ band by compiling from the
literature a sample of nearby dwarf irregular galaxies selected for spanning a wide
range in luminosity, having oxygen abundances obtained following an 
homogeneous method (as far as possible), and having reliable distance determinations.

\subsection {\label{dataused}Data used in our compilation}

 In Table~\ref{dIrr}
we list absolute $K_s$ magnitudes and oxygen abundances for a number
of nearby irregular galaxies.  The sample of 29 galaxies was taken
either from the recent work of \citet{v05}, where new NIR measurements
of dwarf irregular galaxies are presented, or from the classic work of
\citet{rm95} and includes two additional galaxies NGC 1705 and NGC 1156
\citep[taken from ][]{he04}.  It contains only nearby (distance $<$ 8 Mpc)
irregular galaxies, for which distances can either be determined from the
brightest-stars (bs) method, tip-of-the-RGB techniques (rgb) or Cepheids
(cep). For one of the galaxies, the available distance was determined
from the Tully-Fisher relation (tf).  All distances were obtained from
the compilation of \citet{kkhm04}, except for the distance to IC 4662,
which was taken from \citet{lee03}. 
\footnote {We did not include the following
galaxies from the list of \citet{v05}, because we were not able to locate
the corresponding measurements for their metallicities in the literature:
Cas 1, Mb1, Orion Dwarf, UGC 4115, UGC 4998, UGC 5692, UGC 5848, UGC
8508, UGC 5979, NGC 3741, NGC 4163, NGC 4190, and Holmberg IV.  On the
other hand, the dwarf irregular galaxies from the list of \citet{rm95}
Sextans A, Sextans B, LMC, SMC, WLM, Leo A, IC 1613, and NGC 2366 were
not included because no NIR total magnitudes are available for them.} 
NIR magnitudes for the galaxies not included in the sample of \citet{v05}
are from the 2MASS extended source catalog \citep{jar00}, except for
NGC 5408 whose $K_s$ magnitude is taken from \citet{npcf03}.

The values of foreground Galactic extinction in the $K$-band
(A$_K$) for the sample galaxies are listed in column 3 of
Table~\ref{dIrr}. These are taken from NED and were determined
following \citet{sch98}. Exceptions are galaxies NGC 1569 and
IC 10 that are located at low Galactic latitudes, where extinction
values derived from the dust maps of \citet{sch98} are not accurate.
For these galaxies we adopted the Galactic extinction values determined
by \citet{ola01} and \citet{rbb01}, respectively.

 The oxygen abundances, compiled from the literature, were determined in two ways.
 Column 6 of Table~\ref{dIrr}
shows determinations based on electronic temperatures measured from
the [\ion{O}{3}] $\lambda$ 4363 line (T$_{\rm e}$-method), and column 7 shows
determinations from the empirical calibration of \citet[][ P-method]{pil01a,pil01b}.
An exception is NGC 3738 for which only a metallicity value determined
with the method of \citet{ep84} was found.
In a few cases, P-method abundances were not available in the literature, but we
could use published values of emission-line intensity ratios (corrected
for extinction) to calculate
oxygen abundances following \citet{pil01a,pil01b}. In particular  
we used line intensities from \citet{lee03} for IC 10 and NGC 1560,
from \citet{ti05} for Mrk 209, from \citet{ks01} for NGC 4789A,
from \citet{mmo91} and \citet{mho96} for IC 2574 and DDO 50,
and from \citet{mmc94} for NGC 5408.

\subsection {\label{lzdiagram}The K-band L-Z diagram for dwarf irregulars}

Abundances against extinction corrected K$_s$ magnitudes are plotted in
Fig.~\ref{LZ} (Table~\ref{dIrr}, values of column 6 plotted against those of 
column 2, after correction
with extinction values of column 3).  Data for all sample galaxies are
used with T$_{\rm e}$-method abundances, when available (i.e. for all
but two galaxies, see Table~\ref{dIrr}, column 6).
A non-weighted least-squares fit to the data gives the following result:
\begin{equation}
\rm 12+\log(O/H) = (-0.14 \pm 0.02)\times M_{Ks} + (5.55 \pm 0.26),
\end{equation}
with a $\sigma$ of 0.15 dex in $\log$(O/H).

{ To establish whether there is a correlation in the data, we calculated the
Spearman's rank
correlation coefficient, a non-parametric measure of correlation, and found
a value of
$-$0.88. This indicates the presence of a good anti-correlation, whose
level of significance can be
inferred by comparison with published tables \citep{wj03}. Taking into
account the number of L-Z pairs in the sample,
the above correlation coefficient indicates that the hypothesis that the
variables are unrelated is rejected at
the 0.1 per cent level of significance.}
{ Residuals of the fit are
shown in the lower panel of Fig. 1}.

A good correlation is also present if 
we plotted the subsample of 25 dwarf galaxies that have
abundances determined through the P-method 
(not plotted here).
Also in this case the correlation is 
good\footnote{We excluded from the fit galaxy
NGC 6822, whose metallicity was determined with the P-method, and which  
deviates considerably
from the L-Z relation \citep{pil01b}.},
with a
Spearman-rank correlation coefficient of $-$0.81, { and 
same level of significance as above.}
The result of the non-weighted least-squares fit
to the data is not significantly different from the one above:
\begin{equation}
\rm 12+\log(O/H) = (-0.13 \pm 0.02)\times M_{Ks} + (5.73 \pm 0.34),
\end{equation}
with a $\sigma$ of 0.16 dex.

\section{\label{obs} Observations and results for HCG 31}

 We have used relation (1) above, combined with
{ available K-band photometry, from \citep{ls04}, and our}
new medium-resolution spectroscopy of the objects in HCG 31, to
study the nature of the members of this
group, as deduced from their location in the K-band L-Z relation for
normal dwarf galaxies.
Our new data
are described in the following.

\subsection {\label{newdata}New data -- Observations}

New imaging and multi-slit spectroscopic observations of HCG 31 were
done with the GMOS instrument, mounted on the Gemini North telescope, on
August 29 and September 21 of 2003,  respectively.

The imaging consisted of $5\times180$ s exposures in the r$^\prime$ band, 
and $5\times240$ s exposures in the g$^\prime$ band. The filters are
from the SDSS system \citep{fuk96}. The
typical FWHM for point sources was $\sim$ 0\farcs75 in all images.
The observations
were performed in photometric conditions.  Fig.~\ref{image1}
displays the r$^\prime$ image of HCG 31. Fig.~\ref{zoomedimages}
displays zoomed pannels of selected regions, { with 20 surface
brightness contour levels
logarithmically spaced from 17.8 to 24.0 mag arcsec$^{-2}$. The positions
of the spectroscopic slits are also indicated.}

Standard reduction steps were performed with the Gemini package GMOS.
After flat-fielding and cleaning from cosmic-ray events, the final frames were
analyzed with the program SExtractor \citep{ber96}. 

The calibration to the standard SDSS system  was made
with the general zero points and
extinction coefficients provided by the Gemini
observatory\footnote{\texttt{www.gemini.edu/sciops/instruments/gmos/gmosPhotStandards.html}}.
The accuracy of the calibration is claimed to be  within 5\% to 8\%.

Three multi-slit exposures of 960 seconds each were obtained through
a mask with 1.0\arcsec\ slits, using the R400 grating, for a final
resolution of 6.0--6.5 \AA, covering approximately the range 4000 --
8000 \AA. Three additional multi-slit exposures of 1200 seconds each
were obtained through a mask with 1.0\arcsec\ slits, using the B600
grating, for a final resolution of 4.5 \AA, covering approximately the
range 3750 -- 6600 \AA.  { The typical FWHM for point sources, measured
on images taken for identification of the field, was $\sim$ 0\farcs6.
Only a few of the observations were performed in photometric conditions.}

Standard procedures were used to reduce the multi-slit spectra using
tasks within the Gemini {\sc IRAF}\footnote{IRAF is distributed by
the National Optical Astronomy Observatories, which are operated
by the Association of Universities for Research in Astronomy, Inc.,
under cooperative agreement with the National Science Foundation.}
package. Wavelength calibration was done using Cu-Ar comparison-lamp
exposures before and after the exposure on the target. Flux calibration
was done using spectroscopic standard stars obtained in the same night
of the observations.  The blue and red spectra were glued together,
after flux calibration, and are shown in Fig.~\ref{spectra}.

\subsection {\label{measuredproperties}Measured 
properties of the HCG 31 objects}

In columns 2 and 3 of Table \ref{properties}
we list the coordinates of all objects in 
the HCG 31 group identified in the r' image.
We include
a new object which we named ``R'', at
RA = 05$^{\rm h}$ 01$^{\rm m}$ 34$^{\rm s}$ and DEC = $-$04\Deg 12\Min 57\Sec\ (JD2000),
located 2.5 arcmin north of the group central object C,
about 1 arcmin northwest
of object Q
(see Figs. \ref{image1} and \ref{zoomedimages}).
As discussed in Section~\ref{Discussion}, this may be one of the best candidates
for a tidal dwarf galaxy in HCG 31. We only obtained a spectrum for one
of the blobs which constitute region R, namely R1.

The remaining columns of Table~\ref{properties} list 
measurements made by us and other authors
on the properties of the objects in the group, such as magnitudes,
H$\alpha$-luminosities, velocities,  metallicities and colours 
of the group members.
Column 4 lists the K
absolute magnitudes, computed from the K apparent magnitudes 
measured in \citet{ls04}, for an adopted distance
to the group of 54.8 Mpc (values for regions H, Q, and R were measured by us,
see details below),
column 5 lists the
logarithm of the luminosity in H$\alpha$, when available, from \citet{ls04}
or from this work (in the latter case they are not corrected for light
loss from the slit, and are therefore lower limits), columns 6 and 7 
list respectively the heliocentric radial velocities of the objects
(with errors) from this work and from the \ion{H}{1} velocities obtained 
by \citet{vm05}, when available.  
The next four columns of Table~\ref{properties}
list four different determinations of 12 + log (O/H): using the T$_e$
method and the N2 estimator \citep[results from][]{ls04}, in columns 8 
and 9, and using the N2 and O3N2 estimators, derived from our own data,
in columns 10 and 11 { (see the definition of these metallicity
estimators in the following subsection). Columns 12 and 13 list the
measured line ratios used in the determination of the metallicities
of columns 10 and 11. 
Equivalent widths of H$\alpha$ are listed in column 14
and g'-r' colours (measured within an aperture of 2") 
are in the last column of Table~\ref{properties}. }
Details are presented below.

{\it Radial velocities:}  

The spectra of all objects marked in Fig.~\ref{zoomedimages} (with
exception of A, A1 and A2) are shown in
Fig.~\ref{spectra}.  All spectra have
emission lines.
The heliocentric velocities of the observed objects,
derived from the redshifts of the brightest lines, are listed in column
6 of Table~\ref{properties} (the errors are the rms of the individual
line measurements). The \ion{H}{1} velocities at the location of
each corresponding object \citep[from][]{vm05} are given in column 7.
We note that the velocities of all optical regions studied here coincide
with the \ion{H}{1} velocities from the channel maps  within the errors
(except for region C), suggesting a physical association of the objects
with the \ion{H}{1} clouds. The velocities are in agreement with those
derived from long-slit spectroscopy by \citet{rhf90} and \citet{h92},
for the objects measured in common.

{\it Metallicities:} 

Besides the metallicity estimates obtained by \citet{ls04}, 
not available for all objects (see columns 8 and 9 of Table~\ref{properties}), 
we have 
computed values for 12 + log (O/H) using two 
metallicity indicators, from our own measurements of the 
line ratios. The results are listed in 
columns 10 and 11 of Table~\ref{properties}. 

Our first
estimate was obtained with the N2 calibrator, following \citet{pp04},
which is defined as the logarithm of the
[NII]$\lambda$6584/H$\alpha$ ratio. The resulting
metallicities and the measured ratios are listed in columns 10 and 13 of 
Table~\ref{properties} respectively. 
Our values in column 10 can be directly compared with those
of column 9, derived by \citet{ls04}. As can be noted, these 
completely independent measurements are very similar, for the
objects measured in common.

Our second estimate was made using the O3N2 index
\citep{pp04}, based on the logarithm of the 
([OIII]$\lambda$5007/H$\beta$)/([NII]$\lambda$6584/H$\alpha$) ratio.
The resulting metallicities and measured ratios 
are listed in columns 11 to 13 of Table~\ref{properties}. 

The rms scatter in the calibration of these estimators is 
$\sim$ 0.25 - 0.4, which is larger than the internal errors;
we then assume that 0.3 is the one-sigma error
of our measurements.  The metallicities inferred for the
objects are very similar within the errors, the mean value being 
12+log(O/H) = 8.3 for our estimates (with either method).

We point out that the metallicity obtained for object Q is very
uncertain given that the weak emission lines are superposed onto
strong absorption lines, hampering a reliable determination of the
metallicity for this object. We include this object in the tables
but do not plot its magnitude/metallicity in the K-Z relation
of Fig~\ref{LZ}.

{\it Equivalent widths and ages of objects E, F, H and R:}

For objects E1, F1, F2, F3, H1, H2 and R1, where hardly any continuum is 
detected (see Fig.~\ref{spectra}),
we assume that the objects are excited by young stars formed in an
instantaneous starburst and use Starburst99 \citep{leit99},
with solar metallicity and Salpeter IMF, to estimate ages
(obtained from the 
equivalent widths of H$\alpha$, as given in column 14 of 
Table~\ref{properties}).

From a comparison between the
observed H$\alpha$ equivalent widths 
and those produced
by Starburst99, we find ages around 3 Myr for all regions but F3 (the
latter has an approximate age of 6 Myr, while the youngest one, R1, 
has an age of 2.6 Myr).

{\it Spectroscopic properties of Q and G:}

The two galaxies Q and G, besides exhibiting  \ion{H}{2}-region-like
emission lines, also have strong Balmer absorption lines, typical
of spectra dominated by A- and early F-type stars, with H$\delta$
equivalent widths significantly larger than that of normal spiral galaxies.
The equivalent widths (EW) of the [OII]$\lambda$3727 emission line
and the H$\delta$ absorption line are often used to classify galaxy
spectra on the basis of their current/past star formation episodes
\citep[e.g.][]{p99,pw00,ms04}.  Galaxy G, with its EW([\ion{O}{2}]) =
$\sim$ 20 \AA\ and EW(H$\delta$) $\sim$ 8\AA, falls in the category of the
so-called e(a) galaxies \citep{p99}, or A+em galaxies, according to
the notation of \citet{bal99}, with still considerable ongoing star formation.
The spectrum of galaxy Q does not include the [OII]$\lambda$3727 line,
however its weak H$\alpha$ emission line indicates that a low level of
current star formation is still present.  Its spectrum is dominated by
A-type stars, as indicated by the particularly strong Balmer absorption
lines, EW(H$\delta$) $\sim$ 13 \AA, a rather extreme and unusual value. 
Also this galaxy can then be considered of
e(a) type.

{\it Aperture (g'-- r') for all objects and K magnitudes of H, Q and R:}

Aperture magnitudes in the g' and r' bands were
obtained for all studied objects, within an aperture of 2 arcsec,
using the task phot in IRAF.

K magnitudes for objects H, Q and R, which were not given in \citet{ls04}
were derived by us using the program Sextractor (\citet{ber96},
parameter {\it magbest}).  We measured these three objects in unpublished
J images of HCG 31 and assumed a J-K colour of 0.345 for all the objects.
We used J-band images from the archive of the New Technology Telescope of
the European Southern Observatory, obtained with the instrument SOFI.
Nine images of HCG 31 obtained on Nov 3rd/2001 were retrieved, 
of which only five contained
objects Q and R (object H was visible in all nine).  Images in the K
band were also available in the archive but they did not go deep enough
to allow measurements of the objects.  Calibration of the instrumental
photometry was done by using stars common in our fields and in the
2MASS images of the group.  Our final {\it magbest} J magnitudes were
18.6$\pm$0.1, 14.80$\pm$0.04 and 19.7$\pm$0.5 for objects H, Q and R respectively. These values were
corrected for the foreground Galactic extinction given in NED, A$_J$
= 0.046 mag and transformed into K magnitudes using a J-K colour of 0.345
(which is an average of the J-K colour of the other members of the group).
The corresponding K absolute magnitudes for the three objects, 
assuming a distance to the
group of 54.8 Mpc, are listed in column 4 of Table~\ref{properties}.

\subsection {\label{kzrelationh31}The K-Z relation for the TDG candidates of HCG 31}

 We overplot in Fig.~\ref{LZ} the values for the metallicity and
K magnitude (values in columns 8, 9 or 10 against those of
column 4 of Table~\ref{properties})
for the components of HCG 31. The data on the metallicities
come either from \citep{ls04} or from
this work. Those from this work (for objects H and R only)
were derived through the N2 calibrator (column 10 of Table~\ref{properties}).
The only value plotted from column 9 was that for the region A1, for
which no other measurement of metallicity is available.

In Fig. 14 of \citet{ls04}, which plots the B-metallicity relation for
the HCG 31 members, galaxies B, C and G are more than 2 magnitudes
off the line for the B-Z relation for normal dwarf galaxies. In our
corresponding Fig.~\ref{LZ}, using K magnitudes instead, these same
objects are closer to the best line that fits the relation for normal
dwarfs, while still in the high-luminosity side.  On the
other hand, for the fainter objects, only objects H and R stand off the
correlation, although also A1, E2 and F1+F2 show a larger metallicity and/or
a fainter magnitude than the best-line fit.  We  suggest that A, B,
C, G and Q are galaxy members of HCG 31 while the fainter objects are
either tidal debris or tidal dwarf galaxies of the group formed due to
the interaction.  This is further discussed in section~\ref{Discussion}.

\section {\label{Discussion}Discussion}

\subsection{\label{generalproperties}General Properties of HCG 31 members}

Table~\ref{history} summarizes all of the properties of the HCG 31 members
either from this work or gathered from the literature.
This table may be important for future modelling of the group, for
comparison between simulations and observations. As can be noted, HCG
31 has been observed almost in all wavelengths.  The optical morphology
of the galaxies in the group is very disturbed, as is also the optical
and HI kinematics.

{  
A point which is worth noting is the large difference between the
equivalent widths of H$\alpha$ of object F3 (EW = 74 \AA~), as compared
to those for F1 and F2 (1508 \AA~ and 1010 \AA~ respectively). This, in
turn, leads to a large difference in the derived ages of F1+F2 (3 Myrs)
and F3 (6 Myrs).  The colours g'-r' of the three blobs are very similar
and are the bluest colours observed for any object in this group (see
last column of Table~\ref{properties}).  Amram et al. 2004 showed that
the rotation curve is flat through F1 and F2 and the velocity of F3 is
$\sim$ 40 $km~s^{-1}$ higher than that for F1 and F2.  The values for
the metallicities are similar for the three blobs (within the errors)
although F3 tends to have higher metallicity than the other two. F3 is
also the lowest surface brightness object of the three. One might think
that the different equivalent widths and the (small) discontinuity in
the kinematics could be hints that F1+F2 is a distinct object from F3,
but we not not believe this is the case, given the detailed analysis of
the kinematics of the region: we have revisited the Fabry-Perot data
cube of HCG 31 (see Amram et al. 2004 for a description of the data)
and have seen that there clearly exists a continuity in the velocity
field between F1+F2 and F3.  The situation for object E is also similar:
although the equivalent width of H$\alpha$ for E1 (740 \AA~) is much
larger than that for E2 (34 \AA~), analysis of the kinematics done in
Amram et al. (2004) shows that E1+E2 form one single object.}

One new idea put forward in section \ref{measuredproperties} and also
listed in Table~\ref{history}  may deserve some discussion.  Although
the optical spectra taken in this study show  mainly HII-region-like
spectra for the HCG 31 regions (see Fig.~\ref{spectra}), for two of
the galaxies, G and Q, the spectra are typical of e(a)-type galaxies.
The properties of the e(a) class of galaxies have been interpreted by
\citet{p99} as a possible indication of dusty starbursts.  However,
in the spectra of Q and G there are no indications of particularly
high values of internal extinction affecting the emission lines,
since the observed H$\alpha$/H$\beta$ ratios are moderate ($\sim$ 3.4
for Q and 4.0 for G). Hence, the interpratation of these two spectra
remains uncertain. Nevertheless, we suggest that galaxies Q and G have
started reducing their star formation rate and are evolving towards a
post-starburst phase.

\subsection {The tidal tails and the TDG candidates of HCG 31}

HCG 31 is completely embedded within an \ion{H}{1} envelop and it has two
main tails: 1) the southern tail, which contains objects E, H, F and G,
is a narrow and linear optical and H$\alpha$ tidal tail, starting from
galaxy C towards the southeast, ending with galaxy F or
perhaps going even further (to the region of diffuse emission situated between
F and G).  This tail has been the subject of several previous studies
\citep[e.g.][]{a04} and 2) the northern tail, including objects A1, Q and
R (see Fig.~\ref{image1}).  The base of this tail, close to galaxy A,
shows an open configuration, suggesting that the material has moved from
its original plane.  Except for the \ion{H}{1} study of \citet{vm05},
this northern tail has never been studied before.

N-body simulations for compact groups show that stellar tidal tails
are transient features that can be easily destroyed due to multiple
interactions \citep[e.g. ][]{a97,b85}.  Gaseous tails may also have
similar fates.  The frequency of occurrence and length of the tidal tails
in galaxy mergers are strong functions of the encounter geometry and the
merger phase.  Galaxies in the pre-merger phase, where the two galaxies
are still distinct but have gone through the first encounter, are expected
to have well developed tidal features, as seen for HCG 31.  As the final
merger takes place, the tidal tails gradually disappear, with the material
in the tails being accreted back onto the remnant or escaping the system
altogether, eventually forming TDGs \citep[e.g. ][]{mdh98,hvg96}.

\begin{figure*}
\resizebox{15cm}{!}{\includegraphics{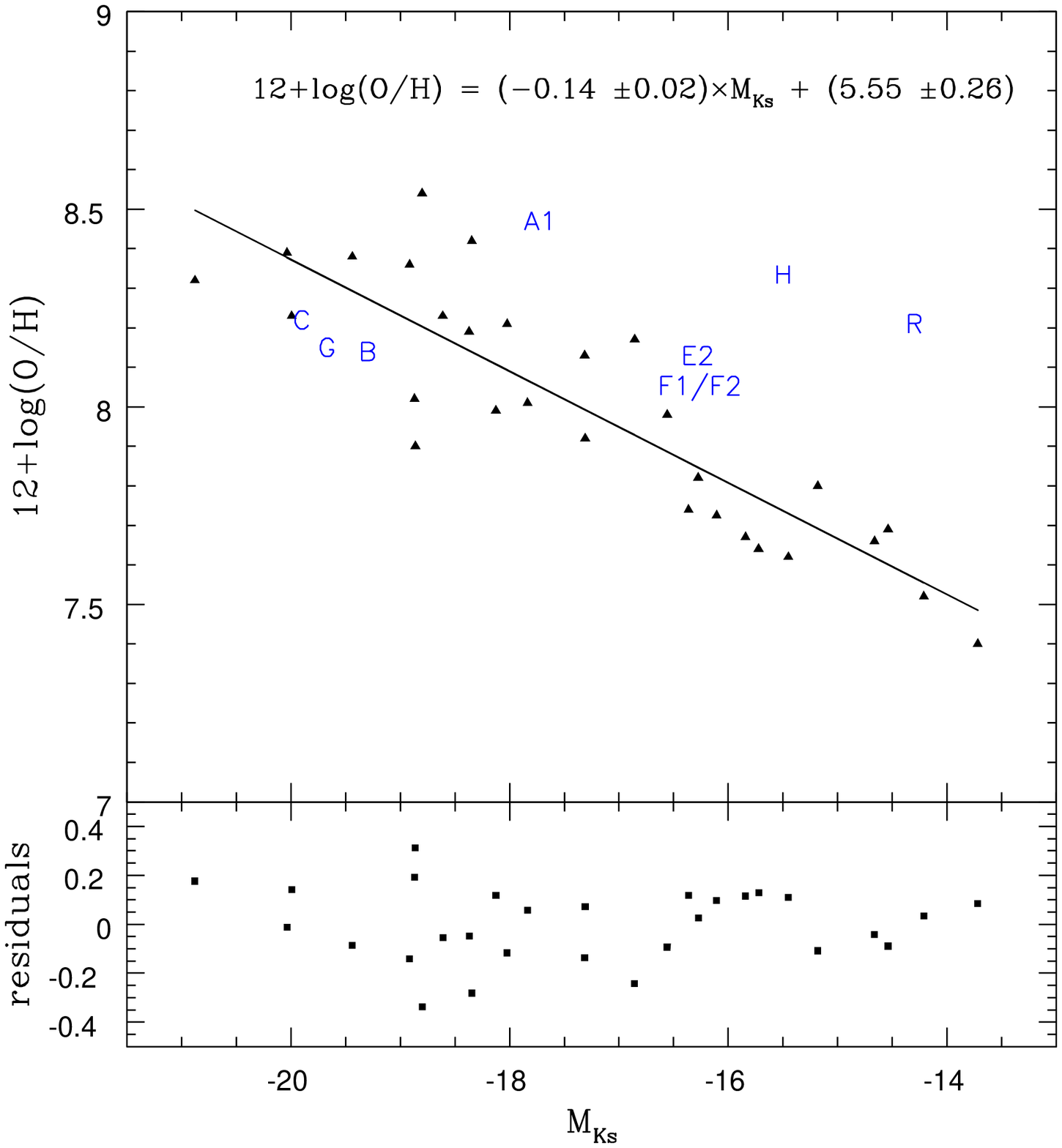}}
\hfill\hfill\caption[]{
$K_s$ L-Z relation for a sample of 29 dwarf irregular galaxies (filled
triangles) out of which 27 have oxygen abundances determined with the
T$_{\rm e}$-method and two have abundances determined from bright emission
lines (Table~\ref{dIrr}).  Overplotted are the components of HCG 31,
indicated by the relevant labels.
Residuals of the fit
are
shown in the lower panel.
\label{LZ} }
\end{figure*}

\begin{figure*}
\resizebox{13cm}{!}{\includegraphics{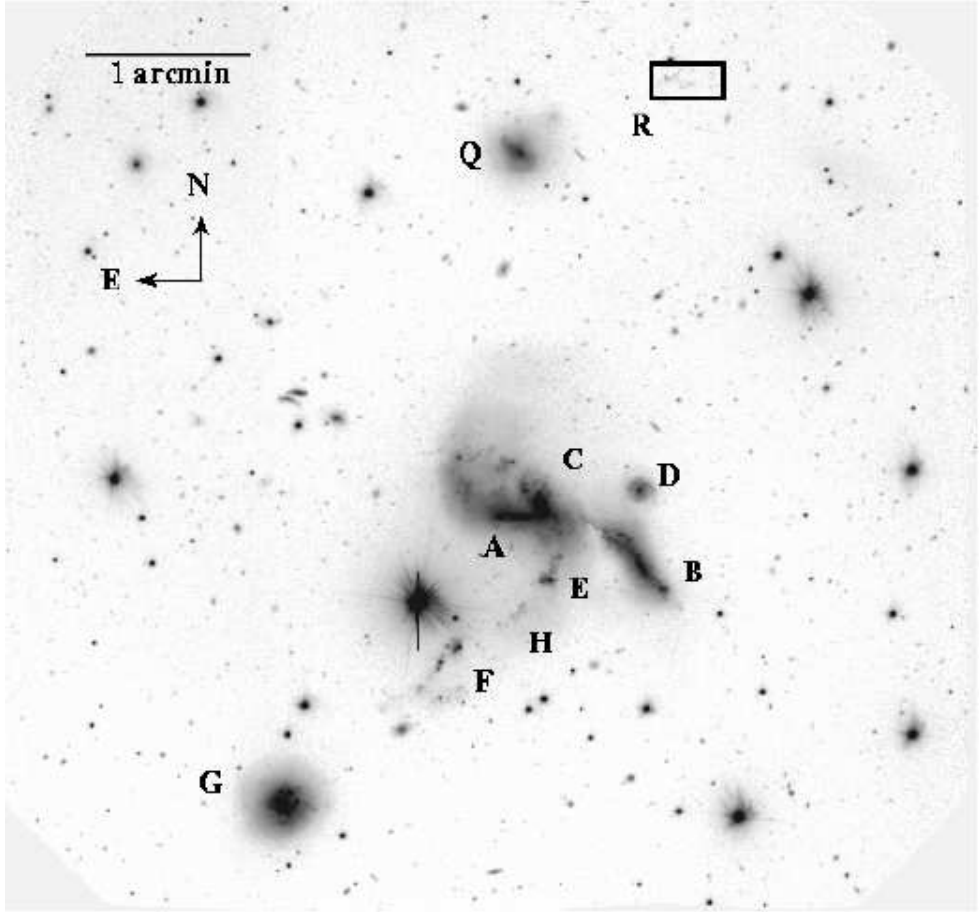}}
\hfill\hfill\caption[]{
Image in the r' band of HCG 31 obtained with Gemini+GMOS, in a
logarithmic gray-scale representation. North
is up and East is to the left. The horizontal bar indicates 1 arcmin.
The individual group members are identified.
All regions marked on the figure belong to the group (i.e. they are at the
same mean redshift of 0.013).
\label{image1}}
\end{figure*}

\begin{figure*}
\resizebox{13cm}{!}{\includegraphics{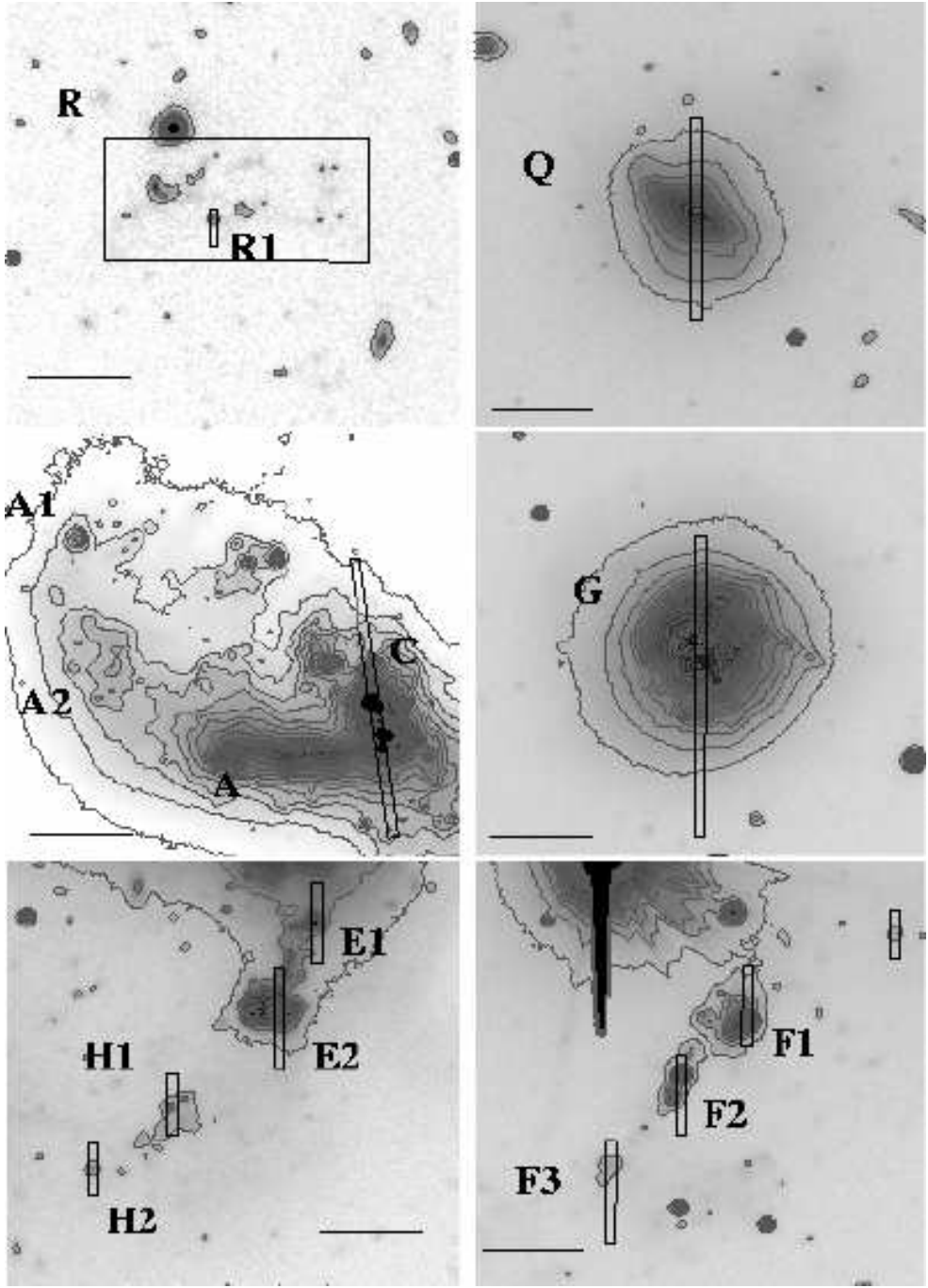}}
\hfill\hfill\caption[]{
Zoomed r' images of various regions of  Fig.~\ref{image1}, to show 
details of the morphology, identification of the components and
placement of the spectroscopic slits. 
The horizontal bar indicates a distance of 10 arcsec. 
{ The 20 surface brightness contour levels 
are spaced logarithmically and go from 
17.8 to 24.0 mag arcsec$^{-2}$. The location and orientation of the
spectroscopic slits are
also shown.}
\label{zoomedimages}}
\end{figure*}

\begin{figure*}
\resizebox{15cm}{!}{\includegraphics{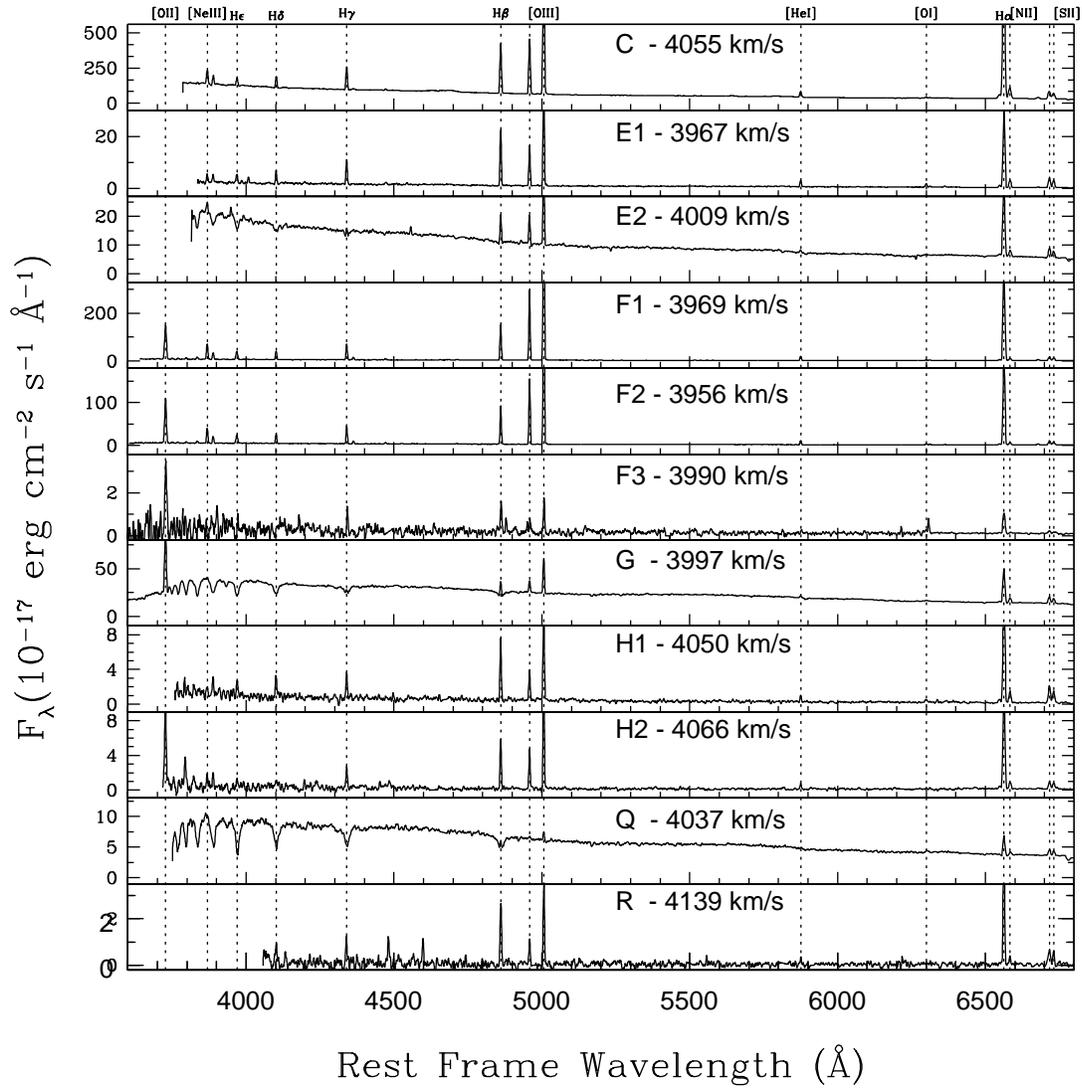}}
\hfill\hfill\caption[]{Spectra of all components of HCG 31 observed
spectroscopically with Gemini/GMOS.
\label{spectra}}
\end{figure*}

In Fig. \ref{LZ} we plotted the location of the components of HCG 31
in the K-Z diagram in an attempt to identify good TDG candidates
in this group.  The objects display a range of luminosities of at least 6
magnitudes and a range of oxygen abundances of $\sim$ 0.5 dex.

While in the B-Z relation plotted in Fig. 14 of \citet{ls04} objects B, C
and G are more than two magnitudes off the relation, in the corresponding
K-Z relation these objects are closer to the best line for normal dwarfs,
although they are still in the high-luminosity side
of the relation.  Although objects A1, E2 and F1/F2 are located in the
low-luminosity side of the relation, they do not stand out, perhaps because
even their K magnitudes could be brightened by the strong star bursts present
in these objects \citep[it is well known that at least F has a low old-stellar 
population content,][]{jc00}.  On the other hand, the low-luminosity
objects H and R are completely off the K-Z relation.
We note that for galaxy Q the
metallicity determination has a very large error given the weak emission
lines in its spectrum -- this is then not plotted in Fig.~\ref{LZ}.

We suggest, based on the velocities,  positions in the L-Z relation, on
their morphologies (our Figs. 2 and 3) and on their internal kinematics
\citep{a04}, that A, B, C, G and Q are galaxy members of HCG 31. On
the other hand, based on the arguments below, we suggest that the
lower-luminosity objects are either tidal debris or tidal dwarf galaxies
of the group, formed as a consequence of the interaction.  The more
difficult task is then to decide which of the lower luminosity objects
A1, E, F, H, R are tidal dwarf candidates or merely tidal debris.

\citet{ipv01} devised a scheme to pick out tidal dwarf galaxy candidates
in compact groups, based on their projected distances to the nuclei
of the parent galaxies (at least 2 R$_{25}$) and their H$\alpha$
luminosities (which should be greater than 10$^{38}$ ergs s$^{-1}$).
We note that HCG 31 was present in the sample of \citet{ipv01} and our
regions E2, H1, H2, F1, F2, and F3 correspond to their regions c, d+e,
f, g, h, and i.  The last three, composing object F, were among their
final list of good candidates for TDGs.
Following the same criteria,
we classify one additional region as a tidal dwarf galaxy candidate,
namely, region R (not known at the time of that study).

\citet{vm05} also noted that object F is a good candidate for a tidal
dwarf in formation.  HCG 31 F1 coincides with a peak \ion{H}{1} column
density of 3 x 10$^{21}$ atoms cm$^{-2}$.  In addition, the three blobs,
F1, F2 and F3 are the bluest objects of the group (see last column of
Table~\ref{properties}).  No underlying old stellar population has been detected for
this object \citep{jc00} and it has a metallicity similar to that of the
central galaxies of HCG 31 (A+C and B).  Although \citet{a04} measured
no rotation for F1+F2, which could suggest that it may not turn into an
independent object, recent simulations of tidal dwarf galaxy formation
\citep{wnb05} conclude that these objects are expected to be non-rotators
or very slow rotators.

Object R is a second excellent candidate tidal dwarf galaxy in
the HCG 31 group. It coincides with an \ion{H}{1} cloud of  column
density of 10$^{21}$ atoms cm$^{-2}$, which is one of the highest seen
among all the tidal filaments of HCG 31. Optically it is
formed by an approximately
round distribution of faint H$\alpha$ knots. We obtained a spectrum
for one of these knots (R1), and we confirm that it is at the redshift of the
group. Different from the intergalactic HII regions observed by \citet{ihii}, 
these knots are located in the peak of the HI distribution and seem to
be a chain of linked star forming regions.

Objects F and R are then
good examples of tidal fragments, well separated from their
progenitors, coincident with peaks of the HI distribution, 
that could evolve to become tidal dwarf galaxies.  
Objects A1, E2 and H
are likely also tidal fragments, but much closer to their
progenitors, and may be falling back to the main central
object of the group. In fact, E has two components in counter-rotation
\citep{a04}, suggesting that some of the material may be
already returning to the parent galaxy.

Our hypothesis that regions F and R are good TDG candidates 
is also supported by the comparison
of their properties with those of tidal dwarf galaxies that have been
identified in other systems, for example those studied by \citet{ipv01},
\citet{wdf03}, and by \citet{twg03}.  The TDGs identified by \citet{ipv01}
in a subsample of HCGs have H$\alpha$ luminosities comparable to those
of the TDG candidates studied in this work.
Similar results concerning H$\alpha$ luminosities
of TDG candidates were obtained by \citet{wdf03}, who found an average
H$\alpha$ luminosity of 2.2 $\times$ 10$^{39}$ ergs s$^{-1}$ for TDGs
identified in a sample of ten interacting systems, with the most luminous
knots having values of order 10$^{40}$ ergs s$^{-1}$.  Other candidate
TDGs were identified by \citet{twg03} in the tidal tail of the compact
group CG J1720-67.8, whose properties seem to indicate an evolutionary
stage similar to that of HCG 31 \citep{tfva05}. The estimated masses
of the TDGs in this system are of order of 2 $\times$ $10^7~M_\odot$
and burst ages for most objects range from $\sim$ 6 to 8 Myr.  H$\alpha$
luminosities from integral field spectroscopy are of order of 10$^{40}$
erg s$^{-1}$, with the brightest TDG candidates having L(H$\alpha$)
= 4 $\times$ 10$^{40}$ erg s$^{-1}$, without correction for internal
extinction \citep{tsk05}.  The above ages are comparable to that obtained
for object F3 (c.f., Section 3.2) and the H$\alpha$ luminosities are
of the same order of magnitude of those of our brightest components,
listed in column 5 of Table~\ref{properties}.  
Additionally, we note that TDGs are
usually associated with \ion{H}{1} density peaks \citep[e.g. TDGs in
M81 and in NGC 5291, ][ respectively]{mak02,dm98}, as it is the case
for objects F and R \citep[see fig. 3 of][]{vm05}.

Regarding the metallicities, when we place the HCG 31 objects in the
diagram of Fig.~\ref{LZ}, we note that for the lowest luminosity
objects (H and R), the metallicites we measure are much higher than those for
normal dwarf irregulars of similar luminosities, although F is not
far from the best line for dwarf galaxies.  We suggest that mostly
the gas and some stars which today form regions F and R were earlier
in galaxies A+C or B and were torned out by the interaction. Then in more
recent times (a few million years ago) the young star complexes in F and
R were formed through compression of the intergalactic \ion{H}{1}
gas by galaxy collisions.  In fact, the young star complexes in R and F
would have had to be formed in situ and not be ejected from the central
galaxy because the time needed for those to move even from the outskirts
of the central interacting pair to their current location, at a typical
ejection velocity (due to dynamical interactions) of 200-300 km s$^{-1}$
would be much longer than the age of their massive stars (as determined
in Section~\ref{measuredproperties}).

N-body simulations of galaxy collisions have shown that tidal dwarf
galaxies often form in the tails of major merger galaxies.  The timing
of each star-forming burst along the tails is strongly determined by
the orbital orientation (prograde or retrograde) and the internal
structure of the merging galaxies (bulge or bulgeless).  Recently,
through high resolution N-body/SPH simulations, \citet{wnb05} have
shown that bound stellar objects can only form in the tidal arms of
interacting disk galaxies, if they have a sufficiently massive and/or
extended gas component.  To our knowledge, no simulation including more
than three galaxies, as is the case for compact groups, has ever been
done, to attempt to reproduce an specific observational configuration.
Simulations of HCG 31 are highly needed to understand the evolution of
this complex interacting system and the formation of new objects.

\acknowledgments

We would like to thank the Gemini staff for obtaining the observations.
The authors would like to acknowledge support from the Brazilian agencies
FAPESP (projeto tem\'atico 01/07342-7), CNPq, DAAD/CAPES (projeto 173/04)
and the Alexander von Humboldt Foundation.  C.M.d.O. and S.T. would like
to thank the hospitality  of the Universitaets-Sternwarte, in Munich and
the Max-Planck-Institut f\"ur Extraterrestrische Physik, in Garching,
where part of this work was developed.  S.T. acknowledges support by the
Austrian Science Fund (FWF) under project P17772.  We made use of the
Hyperleda database and the NASA/IPAC Extragalactic Database (NED). The
latter is operated by the Jet Propulsion Laboratory, California Institute
of Technology, under contract with NASA.

\clearpage

%FIGURE 1 

%FIGURE 2 

%FIGURE 3 

\clearpage

%FIGURE 4 

%__________________________________________
%
%Table 1: K and Z of dwarf Irregulars
%__________________________________________

\begin{deluxetable}{lrrrrrrr}
\tablecolumns{7}
\tablewidth{0pc}
\tablecaption{\sc $K_s$ Magnitudes and Metallicities of Dwarf Irregular Galaxies \label{dIrr}}
\tablehead{
\colhead{Name} & \colhead{M$_{K_s}$} &\colhead{A$_K$} &\colhead{D.M.} & \colhead{M.} & 
\multicolumn{2}{c}{12$+\log$(O/H)} &
\colhead{References}\\
\colhead{} & \colhead{(mag)} &\colhead{(mag)}  &\colhead{(mag)} & \colhead{} & \colhead{(T$_{\rm e}$-method)} & \colhead{(P-method)} & \colhead{}}
\startdata
DDO 47   & -15.17  &  0.012   &28.57  & rgb&  7.80  &  7.86 & 3, 4 \\
DDO 50   & -18.85  & 0.012   &27.65  & rgb&   7.90 & 7.84 & 13 + 19, this work \\
DDO 167  & -14.66  & 0.004   &28.11  & rgb&  7.66  &  7.81 & 14, 4\\
DDO 187  & -14.53  & 0.009   &26.99  & rgb&  7.69  &  7.62 &  7, 4\\
GR 8     & -13.71  &  0.01   &26.61  & rgb&  7.40  &  7.60 &  11, 4\\
IC 10\tablenotemark{a}    & -18.09  &  0.28  &24.10  & cep&   8.19 &  8.25 & 7, this work \\
IC 2574  & -17.30  & 0.013   &28.02  & rgb&   8.13 & 8.22 & 6+19, this work\\
IC 4662\tablenotemark{a}  & -17.81  & 0.026   &27.32  & \nodata  &   8.01 & 8.03\tablenotemark{e} &  13, 4\\
IC 5152  & -17.30  & 0.009   &26.58  & rgb&   7.92 & 7.80 &  12, 12\\
Mrk 178\tablenotemark{a}  & -16.10  & 0.007   &27.95  & rgb&  7.73  &  7.88 & 3, 4 \\
Mrk 209\tablenotemark{a}  & -16.27  & 0.005   &28.78  & bs &  7.82  & 7.76 & 9, this work \\
NGC 55   & -20.03  & 0.005   &26.28  &  tf&   8.39 &  8.29 & 1, 1 \\
NGC 1156 & -19.91  & 0.082   &29.46  & bs &   8.23 &  \nodata & 18 \\
NGC 1560 & -18.80  & 0.069   &27.69  & rgb&   8.02\tablenotemark{b} & 8.12 &  17 , this work\\
NGC 1569 & -18.41  &    0.2   &26.45  & bs &  8.23  &  8.31 & 1, 1 \\
NGC 1705\tablenotemark{a} & -18.02  & 0.003   &28.54  & rgb&   8.21 &  8.13 & 16, 16 \\
NGC 3109 & -16.34  & 0.024   &25.62  & rgb&   7.74 &  7.63 & 7, 12 \\
NGC 3738 & -18.91  & 0.004   &28.45  & rgb&\nodata &  8.36\tablenotemark{c} & 8 \\
NGC 4214 & -19.43  & 0.008   &27.34  & rgb&   8.38 &  8.32 & 1, 1 \\
NGC 4395 & -18.34  & 0.006   &28.32  & rgb&   8.42 &  8.27 &  1, 1\\
NGC 4449 & -20.87  & 0.007   &28.12  & rgb&   8.32 & \nodata & 14 \\
NGC 4789A& -15.84  & 0.003   &28.17  & bs &  7.67  & 7.74 &  10, this work \\
NGC 5264 & -18.78  & 0.019   &28.28  & rgb&\nodata &  8.54\tablenotemark{d} & 12 \\
NGC 5408\tablenotemark{a} & -18.10\tablenotemark{e}  & 0.025   &28.41  & rgb&   7.99 & 7.88\tablenotemark{f} &  15, this work \\
NGC 6822 & -16.77  & 0.087   &23.49  & cep&   8.17 & 8.35 & 4, 4 \\
UGCA 92  & -15.16  &  0.291   &26.28  & bs &  7.62  & \nodata & 2 \\
UGC 4483 & -14.20  &  0.012   &27.53  & rgb&  7.52  &  7.47 & 5, 4 \\
UGC 5423 & -16.53  &  0.029   &28.62  & bs &  7.98  &  7.81 & 6, 4 \\
UGC 6456 & -15.71  &  0.013   &28.19  & rgb&  7.64  &  7.71 & 7, 4\\
\enddata
\tablecomments{Column 1: Identification of the object. Column 2: $K_s$ magnitudes either 
from \citet{v05} or from 2MASS, before correction for Galactic extinction. 
Column 3: Galactic extinction from NED or from \citet[][NGC 1569]{ola01} and \citet[][IC 10]{rbb01}.
Column 4: Distance moduli (D. M.) from \citet{kkhm04}. 
Column 5: Methods used for distance determination:
`brightest stars' (bs), `tip of RGB' (rgb), and Tully-Fisher relation (tf). The
D. M. for IC 4662 is from \citet{lee03}. Columns 6 and 7: Oxygen abundances derived
with the T$_{\rm e}$-method and the P-method, respectively.
When abundances were present in the literature for more than one \ion{H}{2} region of a given galaxy,
the average value was considered. Column 8: References for the metal abundances; 
when two references are indicated, the first is for the T$_{\rm e}$-method and the second for the 
P-method determination.}
\tablenotetext{a}{These galaxies have been classified as blue compact dwarfs by some authors:
\citet[][ IC 10]{ri01}; \citet[][ Mrk 209]{ho99}; \citet[][ Mrk 178, IC 4462, NGC 5408]{npcf03};
\citet[][ NGC 1705]{gmp03}}
\tablenotetext{b}{\citet{lee03} derived for NGC 1560 a lower limit 12$+\log$(O/H) $\geq$ 7.97}
\tablenotetext{c}{Metallicity derived with the empirical method of \citet{ep84}}
\tablenotetext{d}{As noted by \citet{lgh03} the P-method value for this galaxy is uncertain, although in 
agreement with other methods of abundance determination. They adopt as best value an average of the values given
by different methods, 12$+\log$(O/H) = 8.61, and note that this galaxy has an anomalously high metallicity
for its luminosity.}
\tablenotetext{e}{$K_s$ magnitude from \citet{npcf03}}
\tablenotetext{f}{\citet{pil01b} note that the \ion{H}{2} regions of IC 4662 seem to belong to the 
transition zone of the R$_{23}$ - O/H diagram, hence, in strict sense, the P-method equation for
low metallicity \ion{H}{2} regions should not be used. A similar argument holds for NGC 5408.}
\tablerefs{(1) \citet{pil01a}; (2) \citet{m98}; (3) \citet{hgo98}; (4) \citet{pil01b}; (5) \citet{sk94};
(6) \citet{mho96}; (7) \citet{lee03}; (8) \citet{ho99}; (9) \citet{ti05}; (10) \citet{ks01}; 
(11) \citet{hgo02}; (12) \citet{lgh03}; (13) \citet{hgso03} ; (14) \citet{skh89}; (15) \citet{mmc94}; 
(16) \citet{ls05}; (17) \citet{rm95}; (18) \citet{vsc87}; (19) \citet{mmo91}}
\end{deluxetable}

%%%%%%%%%%%%%%%%%%%%%%%%%%%%%%%%%%%%%%%%%%%%%%%%%%%%%%%%%%%%%%%%%%%
%%%%%%%%%%%%%%%%%%%%%%%%%%%%%%%%%%%%%%%%%%%%%%%%%%%%%%%%%%%%%%%%%%%%
%
%Table 2: Properties of the objects in HCG 31
%
%%%%%%%%%%%%%%%%%%%%%%%%%%%%%%%%%%%%%%%%%%%%%%%%%%%%%%%%%%%%%%%%%%%%
%%%%%%%%%%%%%%%%%%%%%%%%%%%%%%%%%%%%%%%%%%%%%%%%%%%%%%%%%%%%%%%%%%%%

\begin{deluxetable}{lcccrcccccccccr}
\tabletypesize{\scriptsize}
\rotate
\setlength{\tabcolsep}{0.05in}
\tablewidth{0pc}
\tablecaption{\sc Properties of the observed objects in HCG 31
\label{properties}}
\tablehead{
\colhead{Name} &
\colhead{$\alpha$} &
\colhead{$\delta$} &
\colhead{M$_K$\tablenotemark{a}} &
\colhead{$\log$ L$_{\rm H\alpha}\tablenotemark{b}-\tablenotemark{c}$} &
\colhead {V$_{Hel}$} &
\colhead {V$_{HI}$} &
\multicolumn{4}{c}{12$+\log$(O/H)} &
\colhead{[OIII]$\lambda$5007}&
\colhead{[NII]$\lambda$6583} &
\colhead{EW (H$\alpha$)} &
\colhead{g$^\prime$ - r$^\prime$} 
\\
\colhead{} &
\multicolumn{2}{c}{(J2000)} &
\colhead{(mag)} &
\colhead{(ergs s$^{-1}$)} &
\colhead{(km s$^{-1}$)} &
\colhead{(km s$^{-1}$)} &
\colhead{T$_e$\tablenotemark{d}} &
\colhead{N2\tablenotemark{d}} &
\colhead{N2} &
\colhead{N2O3} &
\colhead{(H$\beta$)} &
\colhead{(H$\alpha$)} &
\colhead{(\AA)} &
\colhead{} 
}
\startdata
A   &  5:01:38.67 & -4:15:33.6 & \nodata   & \nodata		   ~ & \nodata	    & 4090    & \nodata & \nodata & \nodata & \nodata & \nodata & \nodata & \nodata & 0.67 \\
A1  &  5:01:39.74 & -4:15:12.4 & $-$17.72  & 38.76\tablenotemark{b}~ & \nodata	    & \nodata & \nodata & 8.47    & \nodata & \nodata & \nodata & \nodata & \nodata & 0.54 \\
B   &  5:01:35.25 & -4:15:51.6 & $-$19.30  & 40.22\tablenotemark{b}~ & \nodata	    & 4122    & 8.14	& 8.39    & \nodata & \nodata & \nodata & \nodata & \nodata & 0.60 \\
C   &  5:01:37.76 & -4:15:28.4 & $-$19.90  & 40.57\tablenotemark{c}~ & 4055$\pm$~34  & 3984    & 8.22	& 8.40    & 8.27    & 8.21    & 3.266	& 0.079   & ~281    & 0.42 \\
E1  &  5:01:37.23 & -4:15:48.9 & \nodata  & 40.05--39.15	   ~ & 3967$\pm$~10  & \nodata & 8.13	& 8.36    & 8.31    & 8.30    & 2.095	& 0.092   & ~740    & 0.59 \\
E2  &  5:01:37.48 & -4:15:57.5 & $-$16.28 & 38.90\tablenotemark{c}~ & 4009$\pm$~14  & \nodata & \nodata & \nodata & 8.30    & 8.25    & 2.770	& 0.088   & ~~34    & 0.42 \\
F1  &  5:01:39.71 & -4:16:22.2 & $-$15.97  & 40.12--40.01	   ~ & 3969$\pm$~~9  & 3968    & 8.07	& 8.06    & 8.00    & 7.98    & 5.850	& 0.026   & 1508    & 0.16 \\
F2  &  5:01:40.15 & -4:16:27.5 & $-$14.67  & 39.67--39.74	   ~ & 3956$\pm$~~9  & \nodata & 8.03	& 8.02    & 7.98    & 7.99    & 5.027	& 0.024   & 1010    & 0.19 \\
F3  &  5:01:40.62 & -4:16:36.5 & \nodata   & 37.87\tablenotemark{c}~ & 3990$\pm$~90  & \nodata & \nodata & \nodata & 8.13    & 8.29    & 1.033	& 0.044   & ~~74    & 0.13 \\
G   &  5:01:44.01 & -4:17:19.5 & $-$19.67  & 40.60--39.16	   ~ & 3997$\pm$~~9  & 4005    & 8.15	& 8.41    & 8.44    & 8.35    & 2.325	& 0.155   & ~~22    & 0.53 \\
H1  &  5:01:38.20 & -4:16:07.2 & $-$15.5\tablenotemark{e}  & 38.17--38.64	   ~ & 4050$\pm$~29  & \nodata & \nodata & 8.41    & 8.33    & 8.52    & 0.459	& 0.102   & ~600    & 0.46 \\
H2  &  5:01:38.72 & -4:16:13.5 & \nodata   & 38.59\tablenotemark{c}~ & 4066$\pm$~23  & \nodata & \nodata & \nodata & 8.24    & 8.23    & 2.444	& 0.069   & ~556    & 0.42 \\
Q   &  5:01:38.30 & -4:13:20.9 & $-$19.3  & 38.11\tablenotemark{c}~ & 4037$\pm$125  & 4090    & \nodata & \nodata & 8.51:   & 8.56:   & 0.694	& 0.207   & ~~~6    & 0.56 \\
R1  &  5:01:34.33 & -4:12:56.7 & $-$14.3\tablenotemark{f}  & 38.12\tablenotemark{c}~ & 4139$\pm$300  & \nodata & \nodata & \nodata & 8.19    & 8.30    & 1.231	& 0.057   & 1672    & 0.44 \\

\enddata
\tablecomments{
Columns 6 to 9 list four independent measurements of 12$+\log$(O/H), column
6 from \citet{ls04}, using the T$_e$ method (row 3 of their
Table 5), column 7 using
the NII/H$\alpha$ ratio from
\citet{dtt02} (row 5 of Table 5 of \citet{ls04}). Column 8 lists results
of our measurements using the
NII/H$\alpha$ ratio (listed in column 11)
and column 9 lists those using the N2O3 ratio (column 10 over
column 11),
both indicators defined in \citet{pp04}.}
\tablenotetext{a}{From Table 2 of \citet{ls04}, for an assumed
distance of 54.8 Mpc, except for the measurements for H, Q, and R,
that were estimated by us from an unpublished J-band
image,
corrected for the foreground Galactic extinction given in NED, A$_J$=0.046 mag,
and adopting J-K=0.345.
}
\tablenotetext{b}{From \citet{ls04}, spectroscopic measurement with corrections}
\tablenotetext{c}{Our measurements; these are lower limits because the
luminosities were not corrected for light loss from the slit.}
\tablenotetext{d}{From Table 5 of \citet{ls04}}
\tablenotetext{e}{This value for the magnitude corresponds 
to the whole region H}
\tablenotetext{f}{This value for the magnitude 
corresponds to the whole region R}
\end{deluxetable}

\clearpage

%%%%%%%%%%%%%%%%%%%%%%%%%%%%%%%%%%%%%%%%%%%%%%%%%%%%%%%%%%%%%%%%%%%
%%%%%%%%%%%%%%%%%%%%%%%%%%%%%%%%%%%%%%%%%%%%%%%%%%%%%%%%%%%%%%%%%%%%
%
%Table 3: History of HCG 31 + Literature data for HCG 31 members
%
%%%%%%%%%%%%%%%%%%%%%%%%%%%%%%%%%%%%%%%%%%%%%%%%%%%%%%%%%%%%%%%%%%%%
%%%%%%%%%%%%%%%%%%%%%%%%%%%%%%%%%%%%%%%%%%%%%%%%%%%%%%%%%%%%%%%%%%%%

\begin{deluxetable}{lll}
\rotate
\tablewidth{0pt}
\tabletypesize{\small}
\tablecaption{History and Properties of HCG 31 members from Literature Data \label{history}}
\tablehead{
\colhead{Nature} & \colhead{Properties} & \colhead{References}}
\startdata
\cutinhead{HCG 31 AC}
Galaxies in pre-          & Underlying old stellar population                   & 1 \\
merging phase             & Wolf-Rayet galaxy                                   & 3 \\
                          & SB nucleus moving to post-SB phase                  & 4 \\
		          & Major interaction occurred 400 Myr ago              & 1 \\
                          & Age of central burst 4 - 5 Myr                      & 1 \\
                          & Young massive clusters $\leq$ 10 Myr                & 1, 2\\
                          & Bulk of 7.7 and 14.3 $\mu$m emission                &  4 \\
		          & Low 14.3 $\mu$m to 7.7 $\mu$m flux ratio            & 4\\
			  & Contain bulk of dust of the system                  & 4 \\
			  & Peak of CO and FIR in overlapping region of A and C & 7\\
			  & Double kinematic components                         & 5 \\
			  & Double nucleus                                      & 6, 2 \\
\cutinhead{HCG 31 B}
Galaxy interacting & Detection at 7.7 $\mu$m indicates SF                    & 4\\
with AC            & Warped disk with SF in 3 H$\alpha$ knots                & 4, 11\\
                   & or tidally distorted spiral arms                        & 10 \\
                   & SF over last 10 Myr with peak at $\sim$ 5 Myr           & 2 \\
		   & Secondary peak of SF $\sim$ 10 Myr ago                  & 2 \\
		   & Metallicity similar to AC                               & 8, 12 \\
		   & Optical and \ion{H}{1} emission are lopsided            & 10 \\
		   & Shift between \ion{H}{1} and stellar light in the south & 10 \\
\cutinhead{HCG 31 E}
Tidal debris removed from AC & Young SSCs (1-3 Myr and $\approx$ 10 Myr)     & \\
Material falling back to AC  & with masses 10$^4$ - 10$^5$ M$_{\sun}$        & 2\\
                             & Age of component E1 $\sim$ 3 Myr              & 13 \\
                             & Underlying old stellar population             & 2 \\
                             & Two components in counter-rotation            & 5 \\
			     & Main component counter-rotating with respect to AC & 5 \\
                             & Amplitude of rotation of main component is 25 km s$^{-1}$) & 5\\ 
                             & Eastern component with recent SF              & \\
			     & Western component is older                    & 2\\
                             & Connected to AC by string of SF regions       & 2 \\
			     & Metallicity similar to AC and B               & 12\\
\tablebreak
\cutinhead{HCG 31 H}
 Tidal debris                & Blob of size of $\sim$ 10\arcsec in between E and F & 12 \\
 removed from AC             & Metallicity similar to AC or higher                 & 12 \\
                             & Higher velocity than E and F by 80 km/s             & 12 \\
			     & SF episode with estimated age  3 - 4 Myr            & 12, 13 \\
\cutinhead{HCG 31 F}
TDG candidate                & Several SSCs ($<$ 4 Myr) with masses 10$^4$ - 10$^5$ M$_{\sun}$ & 2\\
                             & No underlying old stellar population                            & 2 \\
                             & Age of components F1 and F2 $\sim$ 3 Myr                        & 13 \\
                             & Age of component F3 $\sim$ 6 Myr                                & 13 \\     
                             & Metallicity similar to AC                                       & 8 \\ 
			     & Formed in tidal debris (4 Myr)                                  &  2 \\
                             & Detected in H$\alpha$ but not at 7.7$\mu$m                      & 4 \\
                             & Two components non-connected by continuum or H$\alpha$          & \\
                             & within the same \ion{H}{1} peak                                 & 2 \\
			     & No rotation, low gaseous velocity dispersion                    & 5\\
			     & \ion{H}{1} kinematically detached from tidal tail               & 10\\
\cutinhead{HCG 31 G}
Galaxy moving through \ion{H}{1} mass    & Kinematically distinct from AC                & 8 \\
or accreting mass from \ion{H}{1} cloud  & Local maximum of common \ion{H}{1} envelope   & 9, 10\\
                                         & Asymmetric H$\alpha$ emission                 & 2 \\ 
					 & SF at border facing southern tail             & 2 \\
                                         & SF over last 10 Myr with peak at $\sim$ 5 Myr & 2 \\
					 & Late-type morphology with very small bulge    & 10 \\
                                         & e(a) or A+em type galaxy -- with considerable ongoing star formation & 13 \\
\cutinhead{HCG 31 A1}
Tidal debris                  & No rotation, low gaseous velocity dispersion & 5\\
in northern tidal tail of AC  & Detected at 7.7$\mu$m                        & 4 \\
\tablebreak
\cutinhead{HCG 31 A2}
Tidal debris removed from AC  & Rotation pattern (amplitude 25 km s$^{-1}$) & 5\\
Material falling back to AC   & Counter-rotating with respect to AC         & 5 \\
                              & Detected at 7.7$\mu$m                       & 4 \\
\cutinhead{HCG 31 Q}
Galaxy                        & Radial velocity compatible with group membership & 8, 10 \\
                              & Local maximum of common \ion{H}{1} envelope      & 9, 10\\
			      & Spectrum dominated by A-type population          & \\
			      & with very strong Balmer absorption lines         & 13 \\
			      & Possibly post-starburst with residual star formation & 13 \\
\cutinhead{HCG 31 R}
Newly identified TDG candidate  & Formed by a number of star forming regions       & 13 \\
                                & Radial velocity compatible with group membership & 13 \\
                                & Age of component R1 $\sim$ 2.6 Myr               & 13 \\
                                & Metallicity similar to AC and B                  & 13 \\
                                & Coincides with a maximum in the \ion{H}{1} envelope & 10, 13 \\
                                & Very weak continuum                            & 12, 13 \\
\enddata
\tablecomments{The following abbreviations are used in this table: FIR (far infrared emission); 
SB (starburst), SF (star formation), SSCs (super-star clusters), W-R (Wolf-Rayet).}
\tablerefs{(1) \citet{jvl99}; (2) \citet{jc00}; (3) \citet{c91}; (4) \citet{oh02};
(5) \citet{a04}; (6) \citet{mb93}; (7) \citet{y97}; (8) \citet{ri03}; (9) \citet{w91}; 
(10) \citet{vm05}; (11) \citet{ipv97}; (12) \citet{ls04}; (13) this work.}
\end{deluxetable}

\end{document}